\documentclass[12pt]{article}

\title{arXiv in the classical Russian literature}

\author{ Z.~K.~Silagadze \\
Budker Institute of Nuclear Physics SB RAS and \\
Novosibirsk State University, 630 090, Novosibirsk, Russia }
\date{}

\begin{document}

\maketitle

{\bf
The creation of arXiv, a highly democratic open access repository of research 
papers in high-energy physics more than twenty years ago \cite{1,2} was 
undoubtedly the most spectacular and landmark  event for the high-energy 
physics community. However, the following curious observation indicates that 
the general cultural significance of this event extends far beyond the 
boundaries assumed at the beginning.
}

Recently {\it Notices of the American Mathematical Society} published \cite{3}
a tribute to Vladimir Arnold, an eminent Russian mathematician who passed away 
in 2010. The tribute contains translated fragments of the Arnold's 1990 
interview which he gave to a Russian magazine Kvant (Quantum). In the 
interview Arnold worries about the fact that the prestige of fundamental 
sciences and of mathematics in particular declined practically in all 
countries. Governments and  modern consumer society are more fond to applied
sciences considering the fundamental research as a largely useless waste of 
money. Arnold considers such an attitude as extremely stupid, much like of the 
attitude of the hog from the I.~A.~Krylov's fable ``The hog under the oak''. 
At the end of the interview, the {\it Notices} exposes an English translation 
of the fable. It ends in the following way (besides the journal, the complete 
text of the translation, along with the Russian original, can be found on the
translator's web page: http://math.berkeley.edu/\verb+~+giventh/ )

\begin{verse}
%                   \hspace*{10mm} The Hog Under the Oak
%
%          A Hog under a mighty Oak \\
%   Had glutted tons of tasty acorns, then, supine, \\
%           Napped in its shade; but when awoke, \\
%   He, with persistence and the snoot of real swine, \\
%           The giant's roots began to undermine. \\
%      "The tree is hurt when they're exposed." \\
%           A Raven on a branch arose. \\
%    "It may dry up and perish - don't you care?" \\
%    "Not in the least!" The Hog raised up its head. \\
%        "Why would the prospect make me scared? \\
%           The tree is useless; be it dead \\
%     Two hundred fifty years, I won't regret a second. \\
%    Nutritious acorns - only that's what's reckoned!" - \\
%      "Ungrateful pig!" the tree exclaimed with scorn.  \\
%        "Had you been fit to turn your mug around  \\
%           You'd have a chance to figure out \\
%           Where your beloved fruit is born." 

      Likewise, an ignoramus in defiance \\
      Is scolding scientists and science, \\
      {\bf And all preprints at lanl\_dot\_gov}, \\
      Oblivious of his partaking fruit thereof. \\
\end{verse}

Krylov wrote this famous fable in 1823. The appearance of arXiv (boldface 
emphasis is mine) in this excellent modernized translation by Alexander 
Givental and Elysee Wilson-Egolf I consider as highly symbolic. At our days 
arXiv is considered as a synonym of wisdom, scientific freedom, enlightenment 
and progress!


\begin{thebibliography}{99}
\bibitem{1}
P.~Ginsparg,
It was twenty years ago today ...,
arXiv:1108.2700 [cs.DL].
%%CITATION = ARXIV:1108.2700;%%

\bibitem{2}
P.~Ginsparg,
arXiv at 20, 
Nature {\bf 476}, 145-147 (2011).
%%CITATION = NATUA,476,145;%%

\bibitem{3}
B.~Khesin and S.~Tabachnikov (coordinating editors),
Tribute to Vladimir Arnold,
Not.\ Amer.\ Math.\ Soc.\  {\bf 59}, 378-399 (2012).
%%CITATION = AMNOA,59,378;%%

\end{thebibliography}
\end{document}